\documentclass[twocolumn,floatfix,superscriptaddress,a4paper,showkeys,nofootinbib]{revtex4-1}
\usepackage[colorlinks=true,linktocpage=true,linkcolor=blue,citecolor=blue,allcolors=blue]{hyperref}
\usepackage{epsfig}
\usepackage{latexsym}
\usepackage[utf8]{inputenc}
\usepackage{xspace}
\usepackage{indentfirst}
\usepackage{enumitem}
\usepackage{color}
\usepackage{placeins}
\usepackage[table,xcdraw]{xcolor}
\usepackage{setspace}
\usepackage{lipsum}
\usepackage{multirow}
\usepackage{hyperref}

\usepackage{graphicx}

\usepackage{todonotes}

\usepackage{amsmath}
\usepackage{amssymb}
\usepackage[english]{babel}
\usepackage{url}
\graphicspath{{figs/}}
\newcommand{\eq}[1]{\begin{align} #1 \end{align}}

\newcommand{\be}{\begin{equation}}
\newcommand{\ee}{\end{equation}}

\newcommand{\mus}{\mu^*}

\begin{document}

\title{Critical point influenced by Bose-Einstein condensation}
\author{V. A. Kuznietsov}
    \affiliation{Bogolyubov Institute for Theoretical Physics, 03680 Kyiv, Ukraine}
    \affiliation{Physics Department, Taras Shevchenko National University of Kyiv, 03022 Kyiv, Ukraine}
\author{O. Savchuk}
   
    \affiliation{GSI Helmholtzzentrum f\"ur Schwerionenforschung GmbH, Planckstr. 1, D-64291 Darmstadt, Germany}

   \affiliation{Bogolyubov Institute for Theoretical Physics, 03680 Kyiv, Ukraine}
    \affiliation{Frankfurt Institute for Advanced Studies, Giersch Science Center, Ruth-Moufang-Str. 1, D-60438 Frankfurt am Main, Germany} 
\author{O. S. Stashko}
\affiliation{Frankfurt Institute for Advanced Studies, Giersch Science Center, Ruth-Moufang-Str. 1, D-60438 Frankfurt am Main, Germany}
 
\author{M. I. Gorenstein}
  \affiliation{Bogolyubov Institute for Theoretical Physics, 03680 Kyiv, Ukraine}
	 \affiliation{Frankfurt Institute for Advanced Studies, Giersch Science Center, Ruth-Moufang-Str. 1, D-60438 Frankfurt am Main, Germany}

\date{\today}

\begin{abstract}
A system of bosons studied within the mean field framework has two fascinating phenomena: a liquid-gas first order phase transition and Bose-Einstein condensation. Interplay between these two phenomena is being investigated. Depending on the mean-field potential parameters one can observe two types of critical points, called "Boltzmann" and "Bose", that belong to different universality classes with distinct sets of critical exponents. 
As examples of Bose and Boltzmann CPs pion and $\alpha$ matter are considered, respectively.
In general, the phase diagram can have 
one of the CPs or both of them.

\end{abstract}
\maketitle

\section{Introduction}
When both attractive and repulsive interactions are present, systems of classical  particles demonstrate  a liquid-gas first order phase transition (FOPT), see, e.g., Ref.~ \cite{LL}. This transition ends with a critical point (CP) 
at temperature 
$T=T_c$ 
where different thermodynamic quantities show abnormal behavior specific for a second order phase transition.
The concept of critical exponents
splits a great diversity of CP phenomena 
into several universality classes with the same critical exponents  (see, e.g., Ref.~\cite{GNS}). 

Bose-Einstein condensation (BEC) in an ideal gas of bosons was predicted many years ago~\cite{Bose:1924mk,einstein1925stizunger}
and  experimentally confirmed for cold atomic gases in magnetic traps
\cite{Anderson198,PhysRevLett.75.1687,PhysRevLett.75.3969,RevModPhys.71.463}. The theory of the BEC phenomenon for interacting particles has been extensively discussed \cite{kapusta_gale_2006,Andersen_2004,griffin1996bose,PhysRevA.88.053633,Watabe_2019}. In particular, modifications of the line of onset of the BEC (further referred to as the BEC line) due to the small repulsive interactions between particles were discussed~\cite{PhysRevLett.83.1703,Baym_2000, Holzmann_1999,Holzmann_2001,PhysRevC.102.035202,Baym_1999,PhysRevLett.83.3770}. Connections between the BEC and percolation phenomena are pointed out in Refs.~\cite{PhysRevE.63.026115,PhysRev.91.1291,PhysRev.91.1301}.

BEC phenomena have been studied in condensed matter physics, nuclear physics, astrophysics, and cosmology (see, e.g., Refs. \cite{Satarov_2017, Begun:2006gj, Begun:2008hq, Strinati_2018, Nozieres:1985zz, PhysRevLett.101.082502, Chavanis:2011cz, Mishustin_2019, Padilla_2019,Harko2022itw}). In most cases, particle interactions need to be taken into account. If the attractive and repulsive forces are present,
then both phenomena -- the  BEC and a liquid-gas FOPT with a CP  -- are simultaneously  expected in systems of interacting bosons. 
An important property of the CP  is infinitely high particle number  fluctuations (see Refs. \cite{Vovchenko2017cbu, Vovchenko2020lju}). Some special features of the mixed liquid-gas phase within the van der Waals model were discussed in Refs. \cite{Satarov2020mp} and \cite{Poberezhnyuk2020cen}.

The phase diagram of strongly interacting matter is one of the most important problems in physics. 
A liquid-gas FOPT  with formation 
of 
the   Bose condensate (BC) in the liquid phase at $T<T_c$
was considered for interacting $\alpha$ particles in Ref.~\cite{Satarov_2017} and for interacting pions in Refs.~\cite{PhysRevC.103.065201} and \cite{Kuz2021}.  
It was also pointed out \cite{Magner2019, Magner2021,Sanzhur2022} that the CP temperature $T_c$ of classical Boltzmann particles increases due to the Bose statistics 
while the Fermi statistics leads  
to the decrease of $T_c$. The CP location changes a by few percent due to the small quantum correction in $\alpha$ matter. On the other hand, in the system of interacting pions the location of a CP found in Refs.~\cite{PhysRevC.103.065201} and \cite{Kuz2021} lies just on the BEC line. Thus, much stronger  Bose effects are expected in the case of pions.
  
In the present paper we investigate the phase diagram of interacting Bose particles within the mean-field model.
The strength of the quantum statistics effects in the vicinity of a CP
depends on the value of the system parameters.
By changing these parameters we find 
two 
qualitatively different scenarios for CP properties, which are denoted as the Boltzmann and Bose CPs.   
The critical exponents for these CPs 
belong to different universality classes. 
For a special region of model parameters, double phase transition with simultaneous presence of the two CPs (Boltzmann and Bose), becomes possible. We find that relativistic effects appear to be crucially important for these double phase transitions.

	The paper is  organized as follows.  The mean-field model results are presented 
	in Sec. \ref{mf}. This section includes the ground state properties, Boltzmann approximation, and effects
	of the Bose statistics. 
	In Sec.~\ref{CPs} the Boltzmann CP and Bose CP are defined. The particle number 
	fluctuations and  the critical exponents are calculated for the Bose CP and compared to those for the Boltzmann CP. Scenarios with double phase transitions are considered in Sec. \ref{double-PT}.
	Summary in Sec. \ref{sec-sum} closes the paper.

\section{mean-field model }\label{mf}
The statistical system in the thermodynamic limit 
is defined in the grand canonical ensemble in terms of the pressure  $p$ as a function of temperature $T$ and chemical potential $\mu $.
All thermodynamic functions can then be found from the pressure function  using thermodynamic identities. 
The  particle number density $n$, entropy density $s$, and energy density $\varepsilon$ are equal to  
\eq{
n \equiv \left(\frac{\partial p}{\partial \mu} \right)_T~,~~
s = \left(\frac{\partial p}{\partial T}\right)_\mu~,~~~
\varepsilon = Ts +\mu n - p~.
} 
For the ideal Bose gas one finds
\eq{
p_{\rm id}(T, \mu) &= \frac{g}{6\pi^2}\int^{\infty}_0 f_k(T, \mu) \frac{k^4 dk}{\sqrt{k^2 + m^2}}~,\label{pid} \\
n_{\rm id}(T, \mu) & \equiv \left(\frac{\partial p_{\rm id}}{\partial \mu} \right)_T= \frac{g}{2\pi^2}\int^{\infty}_0 f_k(T, \mu) k^2 dk~,\label{nid}
\\
f_k(T,\mu) &= \left[\exp{\left(\frac{\sqrt{k^2 + m^2} - \mu}{T}\right) - 1}
\right]^{-1}~, \label{fk}
}
where $m$ and  $g$ are the particle mass and degeneracy factor, 
respectively~\footnote {We use the units with  $\hbar = c = 1$.}~.

In relativistic systems the chemical potential $\mu$ regulates the conserved charge, i.e., the number of particles minus the number of antiparticles.  The thermodynamic functions of antiparticles correspond 
to a substitution of $\mu \rightarrow-~\mu$. In what follows we are mostly interested in the region of large chemical potentials, $\mu/T >  1$. Thus, a contribution from antiparticles to the system thermodynamics can be neglected. The number of particles plays then the role of the conserved charge.  To have closed form expressions we also often exploit the validity of the non-relativistic approximation, $\sqrt{k^2+m^2}\cong m+k^2/(2m)$, accurate  for $T/m \ll 1$.

The mean-field model of interacting bosons is given by the following set of self-consistent equations (see, e.g., Ref.~\cite{Anchishkin_2015}):
\eq{p(T, \mu) &= p_{\rm id}(T, \mus) + \int^n_0 \frac{d U}{d n'} n' dn'~, \label{p}
\\ 
n(T, \mu) &= n_{\rm id}(T, \mus) + n_{\rm BC}~, \label{n}
\\
\mus &= \mu - U(n)~, \label{mu*}
}
where the density dependent mean-field potential $U$ will be taken in the simple form
\eq{U(n) = -An + Bn^2, ~~~A >0, ~~B>0~. \label{Un}
}
The constants $A$ and $B$ correspond to the attractive and repulsive interactions, respectively.
The quantity 
$n_{\rm BC}\ge 0$ in Eq.~(\ref{n}) is the density  of 
the BC. It corresponds to particles at the zero-momentum state $k=0$. Non-zero values $n_{\rm BC}>0$ can only appear at $\mu^*=m$.  Note that values of $\mu^*>m$ are forbidden as they lead to negative values of particle numbers $f_k$ at small $k$.
The relation
\eq{
\mu^*(T,n) ~ =~ m -0 \label{BCL} 
}
defines the BEC line in the $(n,T)$ plane. At this line, one still has $n_{\rm BC}=0$, and it  corresponds to an onset of the BE condensation.  

\subsection{Ground state at $T=0$}
At $T \rightarrow 0$ the thermal pressure \eqref{pid} and particle number density \eqref{nid}  go to zero, so the system can only exist in a form of the BC. The condition of the BEC, $\mu^*  = m$, can be then rewritten as
\eq{Bn^2_{\rm BC} - An_{\rm BC} - (\mu - m) = 0~.\label{cdm}}
This yields the BC density  $n_{\rm BC}$ as a function of the chemical potential $\mu$ 

\eq{n_{\rm BC}(T =0, \mu) & = \frac{A + \sqrt{A^2 + 4B(\mu - m)}}{2B}~~.}

At $T=0$ and
\eq{\label{mu0}
\mu~<~ \mu_{0} = m - \frac{3 A^2}{16 B}
}
the stable thermodynamic solution corresponds to  $n_{\rm BC}=0$ and $p=0$, 
and at $\mu \ge \mu_0$  
the pressure behaves as 
\eq{p  = - \frac{A n^2_{\rm BC}}{2} + \frac{2B n^3_{\rm BC}}{3} \ge  0~. \label{p0}
}
The first-order phase transition for $T=0$ takes place at $\mu=\mu_0$ 
(see, e.g., Ref. \cite{Kuz2021})
with jumps of $n$ and $\varepsilon$ from their zero values to 
\eq{\label{gs}
& n_{\rm BC}(T=0,\mu=\mu_0) = \frac{3A}{4B}~, ~~~ 
\\
& \varepsilon (T=0,\mu=\mu_0) = 
\mu_0 n_{\rm BC} = \frac{3A}{4B}\left(m - \frac{3 
A^2}{16 B}\right)~.
\label{gs-1}
}
The quantities (\ref{gs}) and (\ref{gs-1})  are referred as the ground state values of the considered system.Note that at all values of $\mu$ they are in agreement with the third law of thermodynamics.

\subsection{Boltzmann approximation}
The Boltzmann approximation 
corresponds to a change of the $f_k$ function (\ref{fk}) to
\eq{
f_k(T,\mu)~=~\exp\left(-~ \frac{ \sqrt{k^2+m^2} -\mu}{T}\right)~.
}
The system pressure in the mean-field model in then reduced to a simple analytical form \cite{PhysRevD.101.014015,  PhysRevC.101.035205}, namely
\eq{p(T, n) = T n - \frac{An^2}{2} + \frac{2Bn^3}{3}~.
\label{boltz}
}
The classical equation of state (\ref{boltz}) depends on the interaction parameters $A$ and $B$, but is insensitive to the particle mass $m$ and degeneracy factor $g$.
The statistical system 
exhibits the  liquid-gas FOPT  with a CP defined by the following conditions   

\eq{
\left(\frac{\partial p}{\partial n}\right)_T = 0~,~~~
\left(\frac{\partial^2 p}{\partial n^2}\right)_T = 0~.\label{cp0}
}
From Eqs.(\ref{boltz}) and (\ref{cp0}) one finds CP parameters

\eq{\label{cp1}
T_{\rm c}^0=\frac{A^2}{8B}~, ~~~ n_{\rm c}^{0}= \frac{A}{4B}~,~~~~ p_c^0=\frac{A^3}{96B^2}~. 
}
In the reduced variables
$\tilde{T}=T/T_c^0$, $\tilde{n}=n/n_c^0$, and $\tilde{p}=p/p_c^0$, 
Eq.~(\ref{boltz}) takes a universal form,
\eq{\label{boltz1}
\tilde{p}~=~3~\tilde{T}~\tilde{n}~-~3~\tilde{n}^2~+~\tilde{n}^3~,
}
independent of the  parameters $A$ and $B$.

\begin{figure*}[!t]
\includegraphics[width=0.93\textwidth]{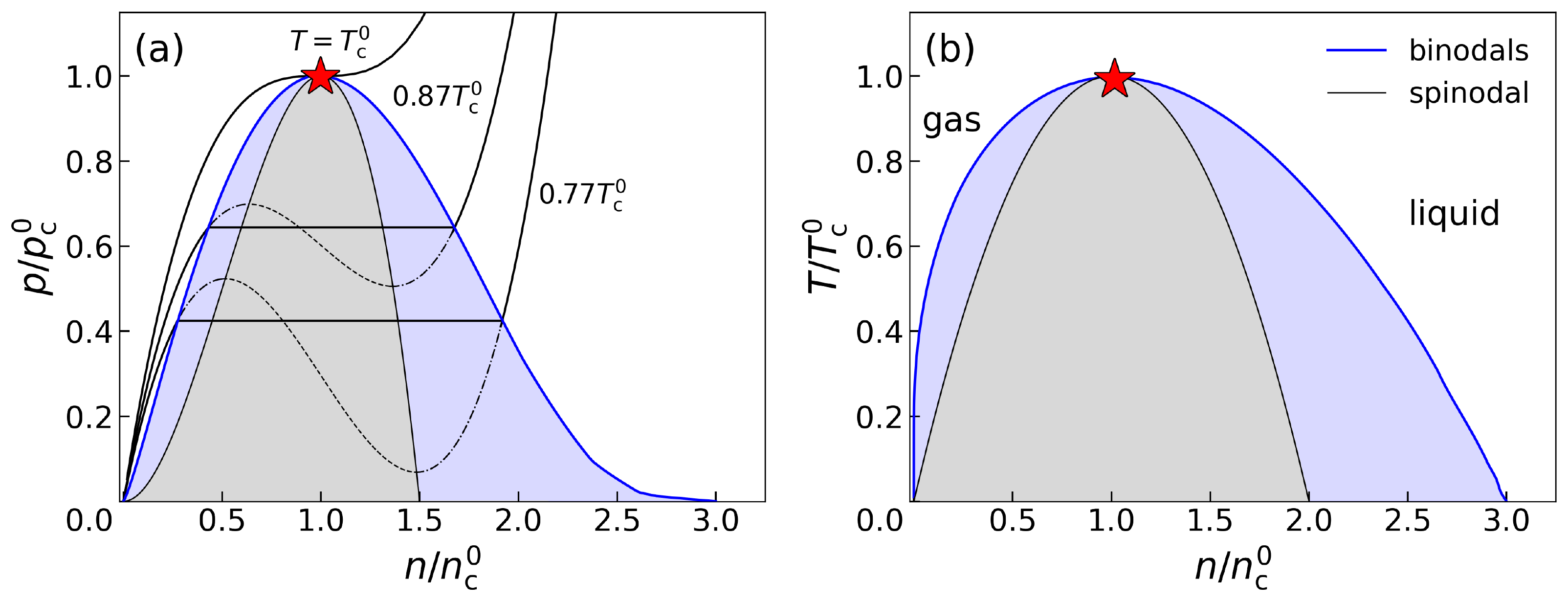}
\caption{\label{fig1}
(a) Phase diagram of the mean-field model in the $(n,P)$ plane within the Boltzmann approximation. Solid lines correspond to the stable parts of the pressure isotherms while dashed and dotted lines show their metastable and unstable  parts, respectively. %
(b) Same phase diagram in the $(n,T)$ plane.}
\end{figure*}

Figure \ref{fig1} demonstrates several pressure isotherms $\tilde{p}=\tilde{p}(\tilde{n})$ at $\tilde{T}\le 1$. 
These isotherms include the metastable and unstable parts denoted in Fig.~\ref{fig1} (a) by  dashed and dotted lines, respectively. A region of the unstable isotherms with $\partial \tilde p(\tilde n,\tilde T)/\partial \tilde n <0 $ shown with grey color in Fig.~\ref{fig1} is restricted by the left and right spinodals where $\partial \tilde p(\tilde n,T)/\partial \tilde n = 0 $. 
In the $(\tilde n,\tilde T)$-regions with $\partial \tilde p(\tilde n,\tilde T)/\partial \tilde n <0$ the system can not exist 
in the homogeneous state. Instead, it behaves as an  inhomogeneous  mixture of the gas and liquid with particle number densities $n_{\rm g}$ and $n_{\rm l}$, respectively. According to the Gibbs criteria these two phases should have the same temperatures, chemical potentials, and  pressures. A region occupied by the liquid-gas mixed phase is restricted  by the left (gaseous) and right (liquid) binodals with particle number densities $n_{\rm g}$ and $n_{\rm l}$, respectively. These binodals are shown by the blue solid lines in Fig.~\ref{fig1} (b). The pressure isotherms inside the mixed phase shown by the horizontal solid lines in Fig.~\ref{fig1} (a)  are obtained from the Gibbs conditions or, equivalently, with the Maxwell construction of equal areas \cite{LL}. 

Critical exponents define the behavior of the thermodynamic function in the vicinity of a CP.
The critical exponents
 $\alpha, \beta, \gamma$, and $\delta$ are defined as:
 \eq{
  \left.\left(\frac{\partial \varepsilon}{\partial T}\right)\right|_{n=n_{\rm c}} ~ & \sim~|T_{\rm c}- T| ^{-\alpha}~, \label{alpha}\\
  n_{\rm l}~-~n_{\rm g} ~& \sim ~\left(T_{\rm c}-T\right)^\beta ~,
 \label{beta} \\
  \left[\left(\frac{\partial p }{\partial n }\right)_{T}\right]^{-1}_{n=n_{\rm c}} ~& \sim ~|T_c-T|^{-\gamma}~, \label{gamma}\\
  \left.\left(p-p_{\rm c}\right)\right|_{T=T_{\rm c}} ~& \sim~|n-n_{\rm c}|^\delta~{\rm sgn}(n-n_{\rm c})~. \label{delta}
 }
 In the Boltzmann approximation (\ref{boltz})  one can easily find 
the following critical exponents in the mean-field model 
 \eq{\label{crexp-boltz}
 \alpha~=~0~,~~~\beta~=~\frac{1}{2}~,~~~\gamma~=~~1~,~~~\delta~=~3~.
 }
 In the present paper we use the Skyrme-like mean-field potential $U(n)$ (\ref{Un}). Its simple form makes possible straightforward  analytical calculations for the first order liquid-gas  phase transition with the end CP for classical particles, i.e., in the absence of the Bose statistics. The CP in the Boltzmann approximation with its critical exponents (\ref{crexp-boltz})  belongs to the universality class of the the so called "classical models" (or the van der Waals type models). This approach gives however only an approximate description of the liquid-gas CP in real systems. 
        A more realistic description of the CP phenomena 
    based on the renormalization group methods  can be found in \cite{Kos2016} and references there in.

The Boltzmann approximation is not valid at very low temperatures $T\rightarrow 0$ and finite particle number density $n$. The Boltzmann gas entropy $s=s(T,n)$ in this limit
becomes negative  in a contradiction with the third law of thermodynamics. Thus the quantum statistics is needed to describe correctly the ground state properties of the system at $T=0$.

\subsection{Effects of Bose statistics}
The thermodynamic properties within considered mean-field model depend on four parameters: $A$, $B$, $m$, and $g$.
The strength  of the Bose effects in a vicinity of a CP  depends on the system parameters. We consider first the small quantum statistics  corrections (see Refs. \cite{Magner2019,Magner2021}). 
The  pressure 
(\ref{p}) 
and particle number density 
(\ref{n}) 
can be presented as 
\eq{\label{p-tot}
& p= 
\alpha~T\sum_{k = 1}^{\infty}K_2\left(\frac{k m}{T}\right)\frac{\exp(k \mu^*/T)}{k^2} -\frac{An^2}{2}+\frac{2Bn^3}{3} , \\
& n = 
\alpha \sum_{k = 1}^{\infty}K_2\left(\frac{km}{T}\right)\frac{\exp(k \mu^*/T)}{k}~+~n_{\rm BC}~, \label{nid-1}
}
where $\alpha\equiv gTm^2/(2\pi^2)$ and $K_2$ is the modified Bessel function. The 
BEC line (\ref{BCL}) obtained from (\ref{nid-1}) at $\mu^*\rightarrow m-0$ and $n_{\rm BC}=0$ reads
\eq{
n ~ = ~
\alpha \sum_{k=1}^\infty K_2\left(\frac{km}{T}\right)~\exp\left(\frac{k~m}{T}\right).
}
Note that in the mean-field model this line in the $(n,T)$ plane is identical to that in the ideal Bose gas.
At low temperatures $T \ll m$, one can use the non-relativistic approximation, i.e., 
$K_2(x) \approx (\pi/(2x))^{1/2} \exp(-x)$,  at  $x\gg 1$. It leads to the well known text-book expression for the BEC line \cite{LL}:
\eq{\label{TBC-1}
T_{\rm BC} ~=~ \frac{2\pi}{m}~\left(\frac{n}{g\zeta(3/2)}\right)^{2/3}~,~~~~\frac{T_{\rm BC}}{m}\ll 1.
}
In the ultra-relativistic limit $T/m \gg 1$, one uses $K_2(x)\approx 2/x^2$ at $x\ll 1$
and finds \cite{Begun:2008hq}
\eq{\label{TBC-2}
T_{\rm BC} =\left[\frac{ \pi^2 n}{g\zeta(3)}\right]^{1/3}~,~~~\frac{T_{\rm BC}}{m}\gg 1~.
}
In Eqs.~(\ref{TBC-1}) and (\ref{TBC-2}), $\zeta(l)$ is the Riemann zeta function, $\zeta(3/2)\approx 2.6 $  and $\zeta(3)\approx 1.2$.

In the non relativistic limit and $z \equiv \exp[(\mu^*-m)/T]\ll1$, one finds 

\eq{
p & \approx gT\left(\frac{mT}{2\pi}\right)^{3/2} \left [z +\frac{z^2}{2^{5/2}}\right]~
-\frac{An^2}{2}+\frac{2Bn^3}{3}~, \label{p-k-2}\\
n & \approx
g \left(\frac{mT}{2\pi}\right)^{3/2} \left [z+\frac{z^2}{2^{3/2}}\right]~.\label{n-k-2}
}
The Boltzmann approximation corresponds to $z\rightarrow 0$ when only the first terms $k=1$ contribute to power series in
Eq.~ (\ref{p-tot}) and (\ref{nid-1}).  

Taking into account the next $k=2$ terms 
one obtains Eqs.~(\ref{p-k-2}) and (\ref{n-k-2}) with corrections due to the Bose statistics that remain  small at $z\ll 1$. 
By inverting $n(z)$ to $z=z(n)$ in Eq.~(\ref{n-k-2}) and substituting it into (\ref{p-k-2}) one finds
the pressure function  
\eq{p(T,n) \approx ~ T n - \left(\frac{A}{2} + \gamma T^{-1/2} \right) n^2 + 
\frac{2B}{3} n^3~, \label{p-app}
}
where 
$\gamma \equiv \pi^{3/2}(2g m^{3/2})^{-1}$. 

The additional $n^2$ term in the pressure (\ref{p-app}) in comparison to its Boltzmann approximation (\ref{boltz}) comes from the (small) effects of the Bose statistics. It corresponds effectively to an increase of the  attractive interactions. Thus, one expects an increase of the CP parameters $n_c^0 \sim A$ and $T_c^0 \sim A^2$. 
Indeed, Eqs.~(\ref{cp0}) for the pressure function (\ref{p-app}) read
\eq{
T - (A + 2 \gamma T^{-1/2} )n + 2Bn^2 &= 0~,\label{cp-1}
\\-(A + 2\gamma T^{-1/2}) + 4Bn &= 0~. \label{cp-2}
}
Solving the system of Eqs. (\ref{cp-1}) and (\ref{cp-2}) for the CP parameters, one obtains
\eq{
n^{1}_{\rm c} & = \frac{n_c^0}{2} \left(~1 ~ + ~\sqrt{1 ~+~ 4 \sigma}~ 
\right) \approx n^0_{\rm c}(1 + \sigma), \label{nc1}\\
T_{\rm c}^1 & ~= ~2B ~(n_{\rm c}^{1})^2~\approx~ T_c^0(1+2\sigma)~ ,
\label{Tc1}
}
where
\eq{\label{sigma}
\sigma~\equiv ~(2 \pi)^{3/2}~\frac {B^{1/2}}{A^2}~ g^{-1}~m^{-3/2}~\equiv~(2\pi)^{3/2} \frac{\tilde{B}^{1/2}}{\tilde{A}^2}~,
}
with the dimensionless parameters 
\eq{\label{AB}
\tilde{A} \equiv gm^2\,A~,~~~~\tilde{B} \equiv g^2m^5\,B~.
}
Note that Eqs.~(\ref{nc1}) and (\ref{Tc1}) are numerically accurate at small $\sigma \ll 1$.  

Equations (\ref{nc1}) and (\ref{Tc1}) demonstrate an increase of $n_c^1$ and $T_c^1$
in comparison with their $n_c^0$ and $T_c^0$ values obtained within the Boltzmann approximation. The size of these Bose effects is regulated by a single parameter $\sigma$ (\ref{sigma}). 
This parameter increases with  the repulsive interactions, $\sigma \sim B^{1/2}$,  and decreases with $A$, $g$, and $m$ as $\sigma \sim A^{-2} g^{-1} m^{-3/2}$. 

To illustrate the role of Bose effects, two examples will be discussed.
The first example concerns a system of interacting $\alpha$ particles ($g=1, m=m_\alpha =3727$~MeV).   Following Ref.~\cite{Satarov_2017} we fix the ground state properties  of the $\alpha$ matter as
\eq{\label{alpha-0}
n_0=\frac{3A}{4B}
=  0.036~{\rm fm}^{-3}~, 
~~W_0=\frac{3A^2}{16B}
= 12~{\rm MeV }.
}
From Eq.~(\ref{alpha-0}) one finds
$A\cong 1.35$~GeV~fm$^{3}$ and $B\cong 28$~GeV~fm$^{6}$. This leads to the CP parameters presented in Table~\ref{table1}.  

\begin{figure*}[t]
\includegraphics[width=0.97\textwidth]{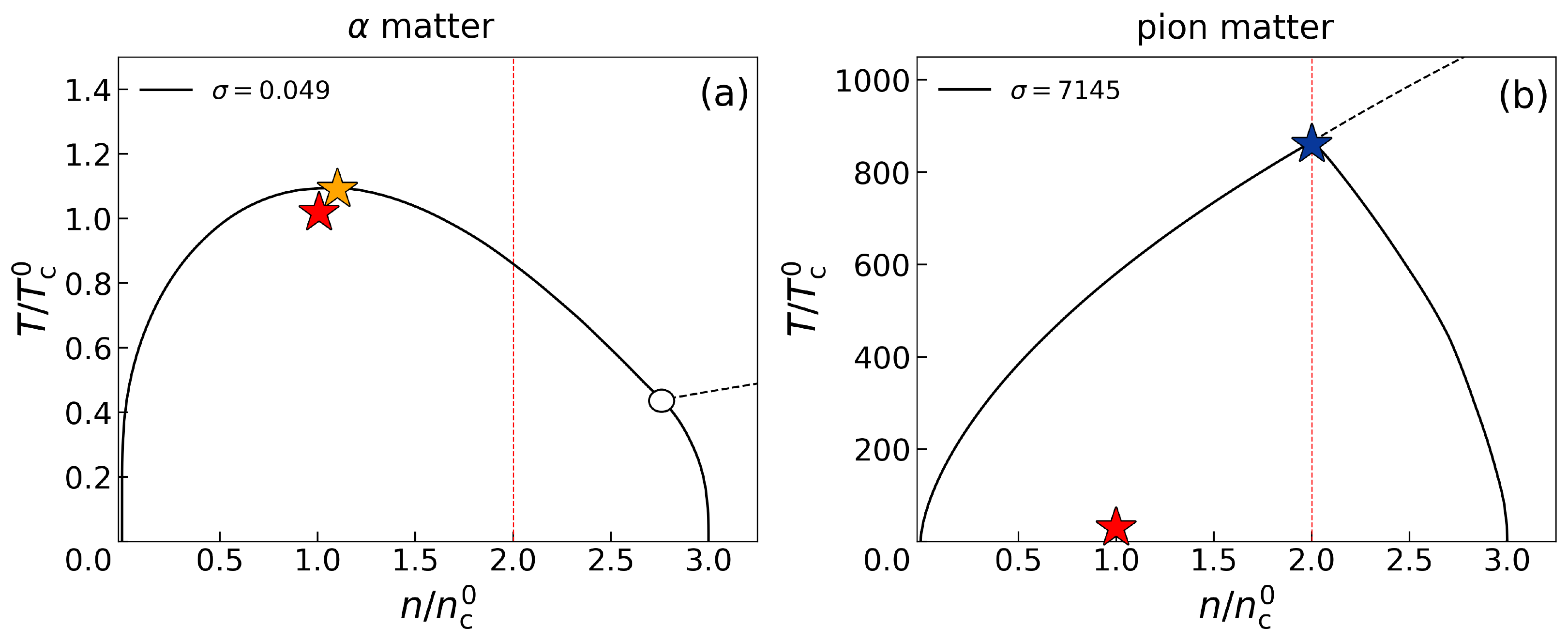}
\caption{The phase diagram of $\alpha$ matter (a) and pion matter (b) in terms 
of the reduced variables $n/n_c^0$ and $T_c/T_c^0$. 
Dashed lines present the BEC lines and solid lines are the gaseous  and liquid binodals. The Boltzmann CP and Bose CP are shown by yellow and blue stars, respectively,  the triple point is shown  by an open  circle,  and red stars correspond to the CP in the Boltzmann approximation.
\label{fig2}} 
\end{figure*}

The phase diagram of $\alpha$ matter is shown in Fig.~\ref{fig2} (a) in terms of the variables $\tilde n= n/n_c^0$ and $\tilde T= T/T_c^0$.
One observes  rather small Bose effects near the CP ($\sigma \approx 0.049  \ll 1$).  

Within the reduced  variables $\tilde{n}$ and $\tilde{T}$ the phase diagram is defined in terms  of the single parameter $\sigma \ll 1$. As seen from Table \ref{table1}, the values of  
$n_{\rm c}^1$ and $T_{\rm c}^1$ obtained in Eqs.~(\ref{nc1}) and (\ref{Tc1}) give a good approximations to their exact values $n_{\rm c}$ and $T_{\rm c}$.

\begin{table}[h!]
\begin{tabular}{|c|c|c|c|c|c|c|}
\hline
$T^{0}_{\rm c}$, MeV & $n^{0}_{\rm c} \ {\rm fm}^{-3}$ & $T^{1}_{\rm c}$ & $n^{1}_{\rm c}$ & $T_{\rm c}$ & $n_{\rm c}$ & $\sigma$ 
\\ \hline
 $8.0 $ &  $0.012 $         &       $8.8$          &     $0.0126$         &     $8.9$         &   $0.013$     &  $0.049$
\\ \hline
\end{tabular}
\caption{Results for the CP parameters for $\alpha$ matter:
$m=3.73$~GeV, $g=1$, $A= 1.35$~GeV~fm$^{3}$, and $B= 28$~GeV~fm$^{6}$.
}
\label{table1}
\end{table}

\begin{table}[h!]
\begin{tabular}{|c|c|c|c|c|}
\hline
$T^{0}_{\rm c}$, MeV & $n^{0}_{\rm c} \ {\rm fm}^{-3}$ &  $T_{\rm c}$ & $n_{\rm c}$ & $\sigma$ 
\\ \hline
 $0.034 $ &  $0.0035 $         &                  $29.2$         &   $0.007$     &  $7\cdot 10^3$
\\ \hline
\end{tabular}
\caption{Results for the CP parameters for the pion matter:
$m=0.14$~GeV, $g=1$, $A=0.0196\ {\rm GeV\cdot fm}^{3}$, and  $\ B=1.426\ {\rm GeV\cdot fm}^{6}$ 
}
\label{table2}
\end{table}

Our second example corresponds to the interacting pion matter ($g=1, m=m_\pi=140$~MeV).
We take  $A=19.6\ {\rm MeV\, fm}^{3}$ and  $\ B=1426\ {\rm MeV\, fm}^{6}$ obtained in Ref.~\cite{Kuz2021} from fitting the lattice data for pion Bose condensate at $T=0$ as a function of electric chemical potential.  The results for $T_{\rm c}$ and
$n_{\rm c}$ values 
are presented in Table \ref{table2}, and the $(n,T)$ phase diagram is shown in Fig.~2 (b). 
In contrast to $\alpha$ matter, the  effects of Bose statistics for pions  are 
very large. One finds $\sigma \gg 1$ and thus $T_{\rm c}/T_{\rm c}^0 \gg 1$. Approximations (\ref{nc1}) and (\ref{Tc1}) lose their validity at large $\sigma \gg 1$,  and thus the $n_{\rm c}^1$ and $T_{\rm c}^1$ values have no physical meaning. 
The BEC line goes through the CP in pion matter, as seen in Fig.~\ref{fig2}(b). 
One finds a limiting value of the critical particle number density,
\eq{\label{nlim}
n_c^{\rm lim} ~=~ \frac{A}{2B}~=~2n_c^0}
that is exactly two times larger than $n_{\rm c}^0$ obtained
in the Boltzmann approximation. Note that the ground state BC density given by Eq.~(\ref{gs}) equals $n_{\rm BC}^0=3n_c^0$.  The limiting critical density $n_c^{\rm lim}$ (\ref{nlim})
corresponds to the $n$ value where $dU/dn=0$, 
and it is independent of parameters $m$ and $g$.

On the other hand, the $T_c$ value at $n_c=n_{\rm c}^{\rm lim}$ is still dependent on $m$ and $g$. One finds
\eq{
T_{\rm c}=T_{\rm BC}~ \approx~ \frac{2\pi}{m}~\left[\frac{n_{\rm c}^{\rm lim}}{g~\zeta(3/2)}\right]^{2/3}~,  ~~~ T/m \ll 1~. 
}

Interacting $\alpha$ matter  and  pion matter are considered within the same mean-field model.  We  are interested in  possible changes of the system parameters $A$, $B$, $m$, and $g$ in the both systems. Our goal  is to study an interplay between the interaction parameters $A$ and $B$, and the Bose statistics effects that are also sensitive to the particle mass $m$ and degeneracy factor $g$.  
In Ref. \cite{Satarov_2017}   the mean-field potential $U(n)$ for $\alpha$ matter was considered as a function of three parameters: $a$, $n_0$, and $\gamma$.   
        In the present study  we use a particular  case of that potential with $\gamma = 1, ~~n_0 = 3A/(4B)$, and $a = A/2$.  This simplifies  the model consideration  and the both  $\alpha$ matter and pion matter are considered within the same model.
 Note also that pure $\alpha$ matter is not a realistic system.  A minimal extension of this model  
     would require one to add the interacting nucleons and some light nuclei  like $d$, $t$, etc. The presence of these additional particles in the nuclear mixture can strongly affects the $\alpha$ particles  \cite{Ropke2009}.  
     We do not insist on the existence of pure $\alpha$ matter in nature.
     The pure $\alpha$ matter in our consideration is only a useful and well studied toy model example to illustrate a limiting case of the Boltzmann-like CP.

\section{Boltzmann and Bose CP regimes}\label{CPs}
The CPs shown in Figs. \ref{fig2}(a) and \ref{fig2}(b) correspond to two limiting cases of small and large effects of quantum statistics. In what follows we denote these two scenarios  as the Boltzmann CP and Bose CP, respectively. 
The reduced variables $n/n^0_{\rm c}$ and $T/T^0_{\rm c}$ appear to be rather useful for the analysis.  
 As discussed in Appendix \ref{A} all possible sets of  the model parameters $A$, $B$, $m$, and $g$ can be combined to the two  dimensionless parameters $\tilde A $ and $\tilde B$ (\ref{AB}) which  define   the type
of the CP. 

In the non-relativistic approximation this is further  simplified to a dependence on the single dimensionless parameter $\sigma$ (\ref{sigma}). 
At fixed  $A$ and $B$ the value of $\sigma$ remains a function of $m$ and $g$. 
One can lead the pion system to a region of small values of $\sigma$ by artificially increasing the pion values of $m$ and/or $g$ 
with fixed $A$ and $B$ values for pions. Similarly one can lead the $\alpha$ system to a region of large values of $\sigma$ by artificially  decreasing of the $m$ value of $\alpha$ with fixed $A$ and $B$ values for $\alpha$ matter. 
These procedures will move the system parameters for pions and $\alpha$ particles between  the regions of small and large values of $\sigma$.

\begin{figure*}[t]
\includegraphics[width=0.97\textwidth]{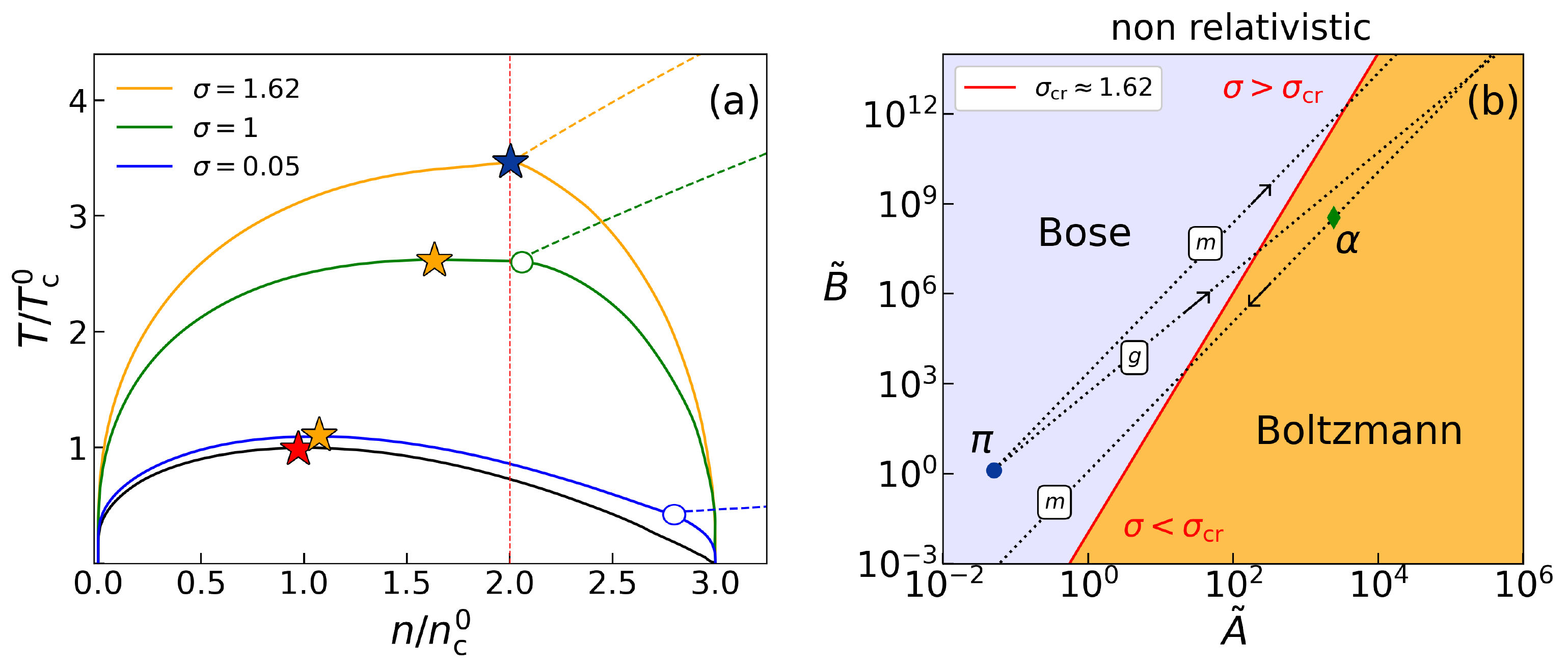}
\caption{(a) Phase diagrams 
at different values of the parameter $\sigma$. 
(b) The $(\tilde A, \tilde B)$ diagram calculated in the non relativistic approximation. The red solid line shows the boundary $\sigma_{\rm cr}=1.62$ between two regimes  (see  Appendix \ref{B}).  Pion matter and alpha matter are indicated with symbols $\pi$ and $\alpha$.
Dotted  lines with arrows illustrate pion matter with increasing $m$ and $g$ values, and $\alpha$ matter with decreasing  $m$ value. }
\label{fig3}
\end{figure*}

The phase boundaries and their CPs for different sets of model parameters obtained by changing the $m$ and $g$ values for pion and $\alpha$ matter
are presented for several values of $\sigma$ 
in Fig.~\ref{fig3}(a).
The  $\sigma$ value work as a single universal parameter.
A common feature for these physical systems is a validity of the non-relativistic approximation, i.e., $T_{\rm c}/m\ll 1$ in all these cases.
A transition between Boltzmann CP and Bose CP regimes takes place at $\sigma=\sigma_{\rm cr}\cong 1.62$. This is discussed in Appendix \ref{B}.

In Figs.~\ref{fig2}(a) and \ref{fig3}(a) open circles denote the triple points, namely the points of intersections of the BEC line with the liquid binodals
at $T_{\rm tr}< T_{\rm c}$. 
The liquid component of the mixed phase at $T<T_{\rm tr}$
includes the BC, $n_{\rm BC}>0$, which affects part of the liquid binodal.    
  In Ref. \cite{Stein1998} a notion of “binodal
anomaly” was introduced to indicate a special form of liquid binodal. It is a result of the influence of the Bose statistics effects. This corresponds to $T_{\rm tr} < T_{\rm c}$ in our notation, in a case when the CP is still the Boltzmann type one. When $T_{\rm tr}=T_{\rm c}$ the BEC affects  the whole  liquid binodal. In addition, this has anomalous influence to the end point of the binodal: the Boltzmann CP is transformed to the Bose CP.  

In Fig.~\ref{fig3}(b) the plane of the parameters $\tilde{A}$ and $\tilde{B}$ is presented. The solid red line  of $\sigma_{\rm cr}\cong 1.62$, defines the boundary between the Boltzmann and Bose regimes. It is calculated in the non-relativistic approximation. The original sets of the $\pi$ and $\alpha$ parameters are shown in the $(\tilde A,\tilde B)$ plane   by the corresponding symbols  in Fig.~\ref{fig3} (b). 
The dotted lines in Fig.~\ref{fig3} (b)  illustrate the changes of the pion system parameters by increasing $m$ or $g$ and the changes of the $\alpha$ system parameters by decreasing $m$. 
For the pion and  $\alpha$ matter parameters $A$ and $B$  the transitions between the Boltzmann and Bose CPs at $\sigma \cong 1.62$ take place at $T_{\rm c}/m \ll 1$, thus, the non-relativistic approximation is valid. 

\subsection{Fluctuations}
The particle number $N$ fluctuates in the grand canonical ensemble. These fluctuations can be characterized by the scaled variance
\eq{\label{omega-1}
\omega~& \equiv ~\frac{\langle N^2 \rangle -\langle N\rangle^2}{\langle N\rangle} ~=~
T~\left[\frac{\partial p(T,n)}{\partial n}\right]_T^{-1}
\nonumber \\
& = \omega_{\rm id}(T,\mu^*)
\left[1 + \frac{n}{T}~\frac{dU}{dn} ~\omega_{\rm id}(T,\mu^*) 
\right]^{-1}~,
}
where 
\eq{
\omega_{\rm id}(T,\mu^*) ~ = ~ \frac{T}{n}~\left(\frac{\partial n}{\partial \mu^*} \right)_T~
}
is the scaled variance for the ideal Bose gas.

In the  Boltzmann approximation, it follows that $\omega_{\rm id}=1$,
and one obtains from (\ref{omega-1}) the simple analytical expression  
\eq{
\omega_0(T,n)= \left[1 + \frac{n}{T}U'(n)\right]^{-1}
= \left[1 - \frac{2\tilde{n}-\tilde{n}^2}{\tilde{T}}\right]^{-1}.
\label{omega0}
}
The gaseous and liquid spinodals in the Boltzmann approximation
shown 
in Fig.~\ref{fig1} are defined by the equation 
\eq{
\left[\frac{\partial \tilde{p}}{\partial{\tilde{n}}}
\right] ~=~\tilde{n}^2 - 2\tilde{n} +\tilde{T} ~=~ 0~.
}
Therefore, along the spinodals $\omega_0 = \infty$, and $\omega_0 <0$ in the unstable $(n,T)$ region between spinodals. 

As seen from Eq.~(\ref{omega-1}), the infinite values of $\omega$ emerge in the Boltzmann CP where
\eq{\label{CPBoltz}
\frac{T_{\rm c}}{n_{\rm c}}~=~ -~~\frac{dU}{dn}~\omega_{\rm id}~,
}
i.e., this requires $dU/dn< 0$. 
In the Boltzmann CP at $n_c<2n_c^0$ the infinite values of $\omega$ 
appear due to the interplay between repulsive and attractive interactions between particles. The presence of attractive interactions leads to
$dU/dn < 0$ at $n=n{\rm _c}<2n_{\rm c}^0$. Equation (\ref{CPBoltz}) 
then leads to 
$\omega\rightarrow \infty$ at the CP $(n_c,T_c)$. This mechanism  does not require Bose statistics, and it takes place already within the Boltzmann approximation, i.e., for $\omega_{\rm id}=1$.  

The infinite fluctuations in the Bose CP at $n_{\rm c} =2n_{\rm c}^0$ have a different origin. On the BEC line one finds \cite{Begun:2008hq}  $\omega_{\rm id}(T,\mu^*\rightarrow m-0)\rightarrow \infty$ and therefore
\eq{\label{CPBose} 
\omega
= \frac{T}{n}~\left(\frac{dU}{dn}\right)^{-1}~.
}
The particle number fluctuations are infinite, $\omega=\infty$, in the Bose CP at $n_{\rm c}=2n_{\rm c}^0$ because of  $dU/dn=0$. At $n> 2n_{\rm c}^0$, $dU/dn>0$ and Eq.~(\ref{CPBose}) lead to finite values of $\omega$ on the BEC line.

\begin{figure*}[t]
\includegraphics[width=0.97\textwidth]{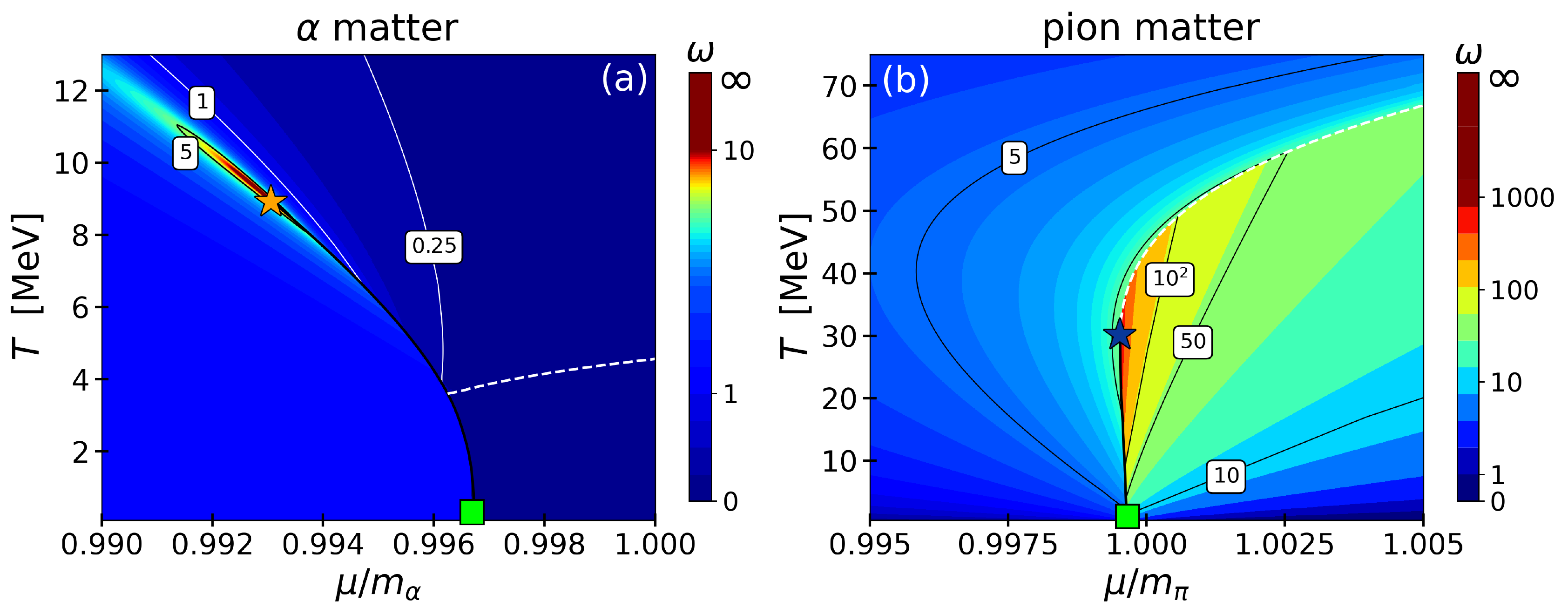}
\caption{The scaled variance $\omega$ in the $(\mu/m,T)$
plane for $\alpha$ matter (a)  and pion matter (b).}
\label{fig-4}
\end{figure*}

The scaled variances $\omega$ for the $\alpha$ matter and pion
matter are shown in Figs.~\ref{fig-4} (a) and (b), respectively, on the $(\mu,T)$ plane. 
They give examples of the Boltzmann and Bose CP. In both cases, infinite values of the scaled variance appear at the CP. However, the regions of large fluctuations look rather different. For the Boltzmann CP shown in Fig.~\ref{fig-4} (a) the large values of $\omega$ are localized in a narrow region of the $(\mu, T)$ plane that looks like a continuation of the FOPT line. The region with the BEC does not generate large fluctuations. For the Bose CP shown in  Fig.~\ref{fig-4} (b) a region of large fluctuations is much wider. It is localized under the BEC line and the FOPT line in the vicinity of the Bose CP.

\subsection{Critical exponents}\label{crexp}
The Bose CP corresponds to $n_{\rm c}=2n_{\rm c}^0$ and
$T_{\rm c}=T_{\rm BC}(n_{\rm c})$. 
The heat capacity (\ref{alpha}) in the considered mean-field model can be calculated in terms of the ideal Bose gas quantities
\eq{\label{alpha-1}
 \left(\frac{\partial \varepsilon}{\partial T}\right)_{n=n_{\rm c}}  \cong  c_{\rm id}(T=T_{\rm c},\mu^*=m )+c( T_{\rm c}- T),~
}
that corresponds to the critical exponent $\alpha=-1$.

At $T<T_{\rm c}$ the chemical and mechanical  equilibrium between the gaseous phase with density $n_{\rm g}$ and the liquid phase with density $n_{\rm l}$  are given by the Gibbs conditions: 
\eq{
\mu_{\rm g}=\mu_{\rm l}~,~~~
p(T,n_{\rm g})=p(T,n_{\rm l})~. \label{gibbs}
}
Equation (\ref{gibbs}) can be rewritten as
\eq{
& m-\mu^*= U(n_{\rm g}) - U(n_{\rm l}) ~,\label{gibbs-1}\\
& p_{\rm id}(T,\mu^*)~-p_{\rm id}(T,m)~=~\int_{n_g}^{n_l}
dn\,n\,dU/dn~.\label{gibbs-2}
}
Using
\eq{
p_{\rm id}(T,\mu^*)~\cong~p_{\rm id}(T,m)~+~n_{\rm id}(T,m)~\left(\mu^*~-~m\right)
}
one finds at small values of $\delta n_{\rm g}= n_{\rm c} - n_{\rm g}$ and $\delta n_{\rm l} =n_{\rm l} - n_{\rm c}$
\eq{
n_{\rm l} -n_{\rm g}\equiv \delta n_{\rm g}+\delta n_{\rm l} \sim (T_c-T)
}
that means $\beta = 1$.
The details are presented in Appendix \ref{C}. In particular, an asymmetry between the gaseous and liquid binodals with $\delta n_{\rm g}/\delta n_{\rm l} \geq 2$ observed. This is different from the properties of the Boltzmann CP where $\delta n_{\rm g}/\delta n_{\rm l}=1$.

Equation (\ref{gamma}) at the Bose CP can be presented as
\eq{
\left[\left(\frac{\partial p }{\partial n }\right)_{T}\right]^{-1}_{n=n_{\rm c}}  =
\frac{\omega}{T} = \frac{ \omega_{\rm id}(T,\mu^*=m))}{T} \sim ~|T_c-T|^{-1} \label{gamma-1}
}
that corresponds to $\gamma=1$, the same as in the Boltzmann CP.

In the vicinity of the  CP at $T=T_{\rm c}$ and $n\rightarrow n_{\rm c}-0$,
 one has $\mu^* \rightarrow m-0$ and finds
 (see also Ref.~\cite{Begun:2008hq})
 \eq{\label{cr-pn}
 & p(T,\mu^*)\cong  p_{\rm id}(T,\mu^*=m)+ n_{\rm id}(T,\mu^*=m)\,
 \left( m-\mu^*\right)\nonumber \\
 &+\int_0^n dn^\prime n^\prime 
 \frac{dU}{dn^\prime} = p_{\rm c} -\int_n^{ n_{\rm c}} dn^\prime n^\prime \frac{dU}{dn^\prime}+ 
n_{\rm c}\, \left(m-\mu^*\right) \nonumber \\
& \cong p_{c}+\left(B-\frac{2\pi^2}{T_c^2g^2 m^{3}}\right)n_c(n-n_c)^2~.
 }
 
 In the non relativistic regime when $T_{\rm c}/m \ll 1 $ the 
 value of $T_{\rm c}/T_{\rm c}^0$ is a function of $\sigma$ given by Eq.~(\ref{TBC-red}). 
 This leads to
 
 \eq{p(T_{\rm c},n) \cong  p_{\rm c} + \left[1 - \left(\frac{\sigma}{\sigma_{\rm cr}}\right)^{2/3}\right]Bn_{\rm c}(n - n_{\rm c})^2~.\label{p-nr}
 }
 
 Equation (\ref{p-nr}) corresponds to the Bose CP in the non-relativistic approximation and, thus, it requires $\sigma> \sigma_{\rm cr}\cong 1.62$.
 In the vicinity of the  CP at $T=T_{\rm c}$ and $n\rightarrow n_{\rm c}+0$,
 one has $\mu^* = m$ and finds
 \eq{p(T_{\rm c},n) \cong  p_{\rm c} + Bn_{\rm c}(n - n_{\rm c})^2~.\label{p-nr-1}
 }
Equations (\ref{p-nr}) and (\ref{p-nr-1}) reveal the critical exponent $\delta=2$ at the Bose CP. Thus, in contrast to the Boltzmann CP, the value of $(\partial^2 p/\partial n^2)_{T=T_{\rm c}}$ at $n=n_c$ is not equal to zero. Instead, this second derivative of the pressure has a discontinuity equal to $A(\sigma/\sigma_{\rm cr})^{2/3}$ at the Bose CP.

To summarize, the set of the critical exponents at the Bose CP,
\eq{\label{cr-exp}
 \alpha~=~-1~,~~~\beta~=1~,~~~\gamma~=~1~,~~~\delta=2~
 }
 is different from that (\ref{crexp-boltz}) at the Boltzmann CP.
 The set of critical exponents (\ref{cr-exp}) corresponds to a universality class of ordinary percolation in $6^+$ dimensions and directed percolation in $4^+$ dimensions (see Refs.~\cite{stauffer2018introduction,SABERI20151,STAUFFER19791}).
Each set of the critical exponents  (\ref{crexp-boltz}) and (\ref{cr-exp}) 
satisfies the scaling relations:
\eq{\label{cexp}
2-\alpha=2\beta+\gamma=\beta(\delta+1)=\gamma\frac{\delta+1}{\delta-1}~.
}

In the special case of $\sigma=\sigma_{\rm cr}$ both sets of the critical exponents (\ref{crexp-boltz}) and (\ref{cr-exp}) are simultaneously realized. The Boltzmann set (\ref{crexp-boltz}) corresponds to $n<n_{\rm c}= 2n_{\rm c}^0$ and the Bose set (\ref{cr-exp}) to $n>n_{\rm c}=2n_{\rm c}^0$. This is a new feature of the critical exponents. For this single value $\sigma=\sigma_{\rm cr}$ the CP keeps a memory of the Boltzmann like behavior at $n<n_{\rm c} $ and the Bose one at $n>n_{\rm c}$.

\section{double phase transitions}\label{double-PT}
Our discussion up to now in most cases assumes the non relativistic approximation $T_{\rm c}^0/m\ll 1$. 
If the relativistic effects become important, another interesting possibility with the two CPs emerges: the Boltzmann CP at $n_{\rm c}< 2n_{\rm c}^0$ 
and the Bose CP at $n_{\rm c}=2n_{\rm c}^0$.
Both conditions (\ref{CPBoltz}) and (\ref{CPBose}) are then simultaneously satisfied, and at the both CPs  the particle number fluctuations are singular with $\omega\rightarrow \infty$. 
For very large attractive interactions such a possibility was observed at zero chemical potential in  Ref. \cite{Stashko2021}. The double phase transitions  at $\mu>0$ will be discussed in this section.

Let us consider first the small Bose effects in the ultrarelativistic limit $m/T \ll 1$.  This permits us to obtain simple analytical approximations.  Taking
into account only the $k=1$ and  $k=2$ terms in the power series (\ref{p-tot}) and (\ref{nid-1}) one obtains   
in the $m/T\ll 1$ limit
\eq{
p_{\rm id}\approx \frac{gT^4}{\pi^2} \left [X+\frac{X^2}{4} \right]~,~~~
n \approx
\frac{gT^3}{\pi^2} \left [X+\frac{X^2}{2} \right]~,
}
where $X=\exp[\mu^*/T]$.
By inverting $n(X)$ to $X=X(n)$ and substituting it into $p_{\rm id}(X)$ one finds
\eq{\label{pid-2}
p_{\rm id} ~ \approx ~ T n ~  - ~ \frac{3\pi^2}{4g T^2} ~n^2~, 
}
and then
\eq{p(T,n) \approx ~ T n - \left(\frac{A}{2} + \frac{\Gamma}{T^2}  \right) n^2 + 
\frac{2B}{3} n^3~. \label{p-rel}
}
where
$\Gamma = 3\pi^2/(4g).$
This is the ultrarelativistic analog of non relativistic approximation (\ref{p-app}).
From Eq.~(\ref{p-rel}) one finds condition for the  CP parameters
\eq{
    T - (A + 2\Gamma T^{-2})n + 2B n^2 &= 0,\label{p1}\\
    - (A + 2\Gamma T^{-2}) + 4Bn &= 0.\label{p2}
}
From Eqs.~(\ref{p1}) and ({\ref{p2}}) one obtains approximate relations
\eq{
n_c^1\approx n_c^0(1 + \kappa), ~~T_c^1 \approx  T_c^0(1 + 2\kappa), ~~
}
valid at
\eq{
\kappa = 
\frac{96\pi^2\tilde B^2}{\tilde A^5} \ll 1~, ~~~~
\frac{m}{T_{\rm c}^0} \ll 1~.
}
With increasing $\kappa$ the critical density $n_{\rm c}$ reaches its limiting value of  $n_{\rm c}^{\rm lim}=2n_{\rm c}^0$,
and the critical temperature behaves as  
\eq{\label{TBC}
T_{\rm c}=T_{\rm BC}~ \approx \left[\frac{\pi^2~n_{\rm c}^{\rm lim}}{g \zeta(3)}\right]^{1/3}~. 
}

Let us denote by $n_{\rm tr}$ 
the particle number density at the triple point, i.e., the point of an intersection of the BEC line with a liquid binodal. This point is shown by an open circle in Fig.~\ref{fig2} (a). 
The values of $n_{\rm c}$ and $n_{\rm tr}$ satisfy the following inequalities:
\eq{
n_{\rm c}^0~<~ n_{\rm c}~\le~ n^{\rm lim}=2n_{\rm c}^0~\le ~n_{\rm tr}~<~3n_{\rm c}^0~.
}

\begin{figure*}[t!]
\includegraphics[width=0.8\textwidth]{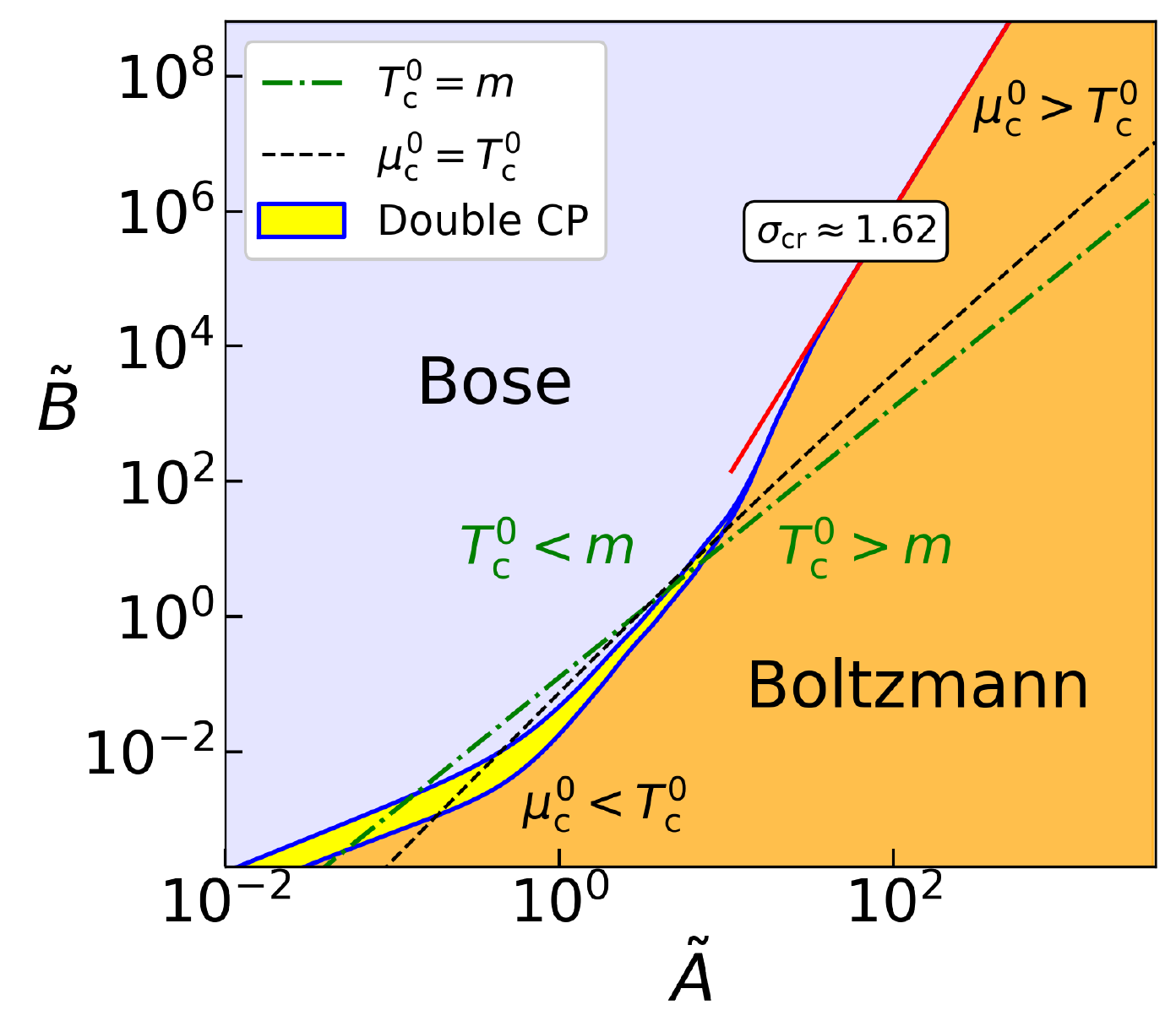}
\caption{\label{fig6} The relativistic $(\tilde A, \tilde B)$ diagram in the mean-field model. The region of the double CPs is shown by yellow color
between the Bose and Boltzmann regions. See text for details.} 
\end{figure*}

\begin{figure*}[!t]
    \includegraphics[width=0.97\textwidth]{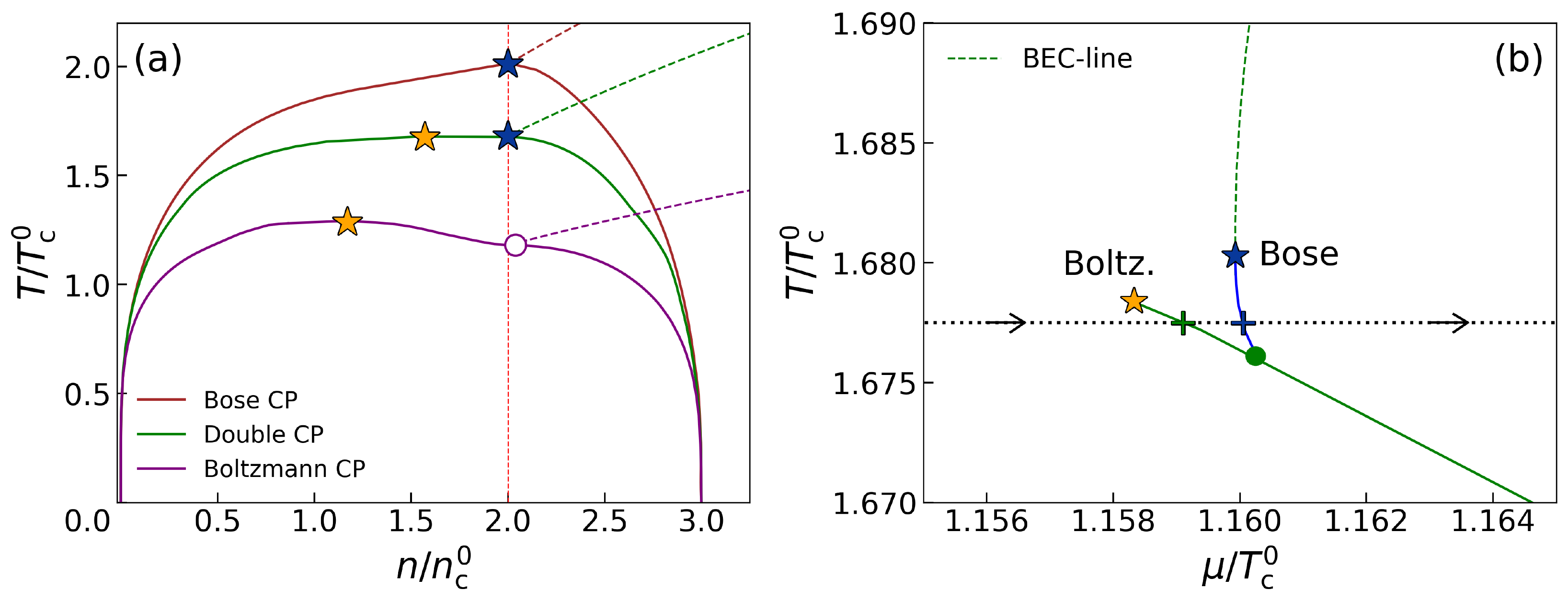}
    \caption{\label{dpt}
    (a) An  example of the phase transitions with two CPs in the $(n,T)$ plane. It corresponds to 
    $\tilde A=12.5$ and $\tilde B= 60$. 
    The Boltzmann CP corresponds to $\tilde A= 8$ and $\tilde B=\ 1.3$,
     and the Bose CP
    to
    $\tilde A=9.7$ and $\tilde B= 59$. 
    (b) The $(\mu,T)$ plane with the double phase transition. Yellow and blue stars correspond to the Boltzmann CP and Bose CP, respectively. 
    A horizontal dotted line illustrates two serial transitions with increasing $\mu$ at fixed $T$.
    }
    \label{fig5}
\end{figure*}

In the non-relativistic approximation a transition to the Bose CP takes place as a merger of the Boltzmann CP and the triple point at $n=2n_{\rm c}^0$ that happens at $\sigma=\sigma_{\rm cr}\cong 1.62$ (see Appendix \ref{B}). 
In a presence of the relativistic effects a transition of the Boltzmann CP to the Bose CP takes place through a double phase transition. This happens in a region of the parameters $\tilde A$ and $\tilde B$ noted as "double CP" in Fig.~\ref{fig6} and shown with yellow color.
Several examples of the simultaneous presence of the two CPs are shown in Fig.~\ref{dpt}. 
A transition from the Boltzmann CP to the Bose CP has the following  general features seen in Fig.~\ref{dpt}. First, the triple point reaches the limiting density $2n_{\rm c}^0$. Thus,
 both the Boltzmann CP at $n=n_{\rm c} < 2n_{\rm c}^0$
and the Bose CP at $n_{\rm c}=2n_{\rm c}^0$ simultaneously exist. Then the Boltzmann CP continues to move to larger $n_{\rm c}$ values, and finally disappears at $n_{\rm c} < 2n^0_{\rm c}$. 

A big variety of the mutual arrangements of the both CP temperatures are found at different sets of the $\tilde A$ and $\tilde B$ values. As seen from Fig.~\ref{fig6}, this $(\tilde A,\tilde B)$ region is not far away from the $T_{\rm c}^0=m$ line, i.e., a presence of the double CPs requires moderate relativistic effects, and it is absent in the both non-relativistic and ultra-relativistic limits.

The double phase  transitions  require  a "quadruple point" where the two phase transition lines in the $(\mu,T)$ plane intersect. 
This point at $T=T_{qd}$ is shown as a full circle in Fig.~\ref{dpt}(b) where 
an example of a double phase transitions is presented in the $(\mu,T)$ phase diagram. The horizontal dotted line in Fig.~\ref{dpt}(b) corresponds a fixed temperature $T$ that is larger than $T_{\rm qd}$
and smaller than both $T_{\rm c}({\rm Boltz})$
and $T_{\rm c}({\rm Bose})$. With increasing of $\mu$ at fixed $T$ the system moves right along the dotted line in Fig. \ref{dpt}(b). Two consecutive phase transitions take place. The first one is between the gaseous and liquid phases, where both phases do not include the BC. The second transition takes place between the liquid phase without the BC and the liquid with the non zero BC density $n_{\rm BC}>0$. 

Note that a contribution of antiparticles to the thermodynamic functions is small in comparison with that from particles at $\mu/T > 1$. In the  Boltzmann approximation
one finds for the CP values
\eq{
\frac{\mu_{\rm c}^0}{T_{\rm c}^0}~=~\ln \left[4\pi^2\tilde A^{-1}~K_2^{-1}(2\tilde B/\tilde A^2)\right]~-~1~.
}
The line $\mu_{\rm c}^0/T_{\rm c}^0 =1$ is shown in the $(\tilde A,\tilde B)$ diagram in Fig. \ref{fig6} by a dashed line. 
From Fig.~\ref{fig6} one observes
that  both relations, 
 $T_{\rm c}^0\approx m $ and $\mu_{\rm c}^0\approx m$, are approximately satisfied in the region of the $(\tilde A,\tilde B)$ parameters  with   the double CP.

\section{Summary}\label{sec-sum}
Thermodynamic properties of interacting Bose particles 
are studied with a relativistic mean-field model. The mean-filed potential is chosen to be $U(n) = -An + Bn^2$, where $A>0$  and $B>0$ terms correspond to attractive and repulsion interactions, respectively. 
The whole range of the system parameters $(A,B,m,g)$
is considered. 
The CP temperature $T_{\rm c}^0$ and particle number density $n_{\rm c}^0$ in the Boltzmann approximation are given by the functions (\ref{cp1}) of the interaction parameters $A$ and $B$, and they are independent of the values of $m$ and $g$. 
The Bose effects are sensitive to the values of $m$ and $g$, and they lead to increase of the $T_{\rm c}$ and $n_{\rm c}$ values in comparison to their $T_{\rm c}^0$ and $n_{\rm c}^0$ Boltzmann limits.

We found two different implementations of the CP, denoted as the Boltzmann CP and Bose CP,  in the system of interacting bosons. These two scenarios correspond to small and large Bose effects, respectively.   
The diagram of the dimensionless parameters 
$(\tilde A,\tilde B)$ 
gives us a possibility to study a diverse physical systems of Bose particles. The interacting  $\alpha$ particles and pions are presented as illustrative examples of the Botzmann CP and Bose CP, respectively. 
The Bose CP has the limiting density $n_{\rm c}=2n_{\rm c}^0$ and the BEC line touches the Bose CP. 
The Boltzmann CP and Bose CP have rather different properties. They have  two different sets of the critical exponents and, thus, belong to distinct  universality classes.
 
On the $(\tilde A,\tilde B)$ diagram a special region of the system parameters was found. In this region   $T_{\rm c}^0 \approx m$ and $\mu_{\rm c}^0 \approx m$ corresponding to the double phase transitions with two CPs -- the Boltzmann CP at $n_{\rm c}< 2n_{\rm c}^0$ and the Bose CP at $n_{\rm c}=2n_{\rm c}^0$.

	\begin{acknowledgments}
	The authors thank M.~Gazdzicki, R.~Poberezhnyuk, L.~Satarov, H.~Stoecker, and V.~Vovchenko for fruitful comments and discussions. This work is supported by the National Academy of Sciences of Ukraine, Grant No. 0121U112254. O.S.S. acknowledges support from National Research Foundation of Ukraine (Project No. 2020.02/0073). M.I.G. acknowledges support from the Alexander von Humboldt Foundation. 
	\end{acknowledgments}

\appendix
\section{Dimensional analysis }
\label{A}
Introducing dimensionless variables
\eq{\hat T = \frac{T}{m}, ~~\hat \mu^*= \frac{\mu^*}{m}~,~~ \hat n = \frac{n}{gm^3}, ~~\hat p = \frac{p}{gm^4} \label{hft}
}
one can rewrite Eqs.~(\ref{p-tot}) and (\ref{nid-1}) as the following functions
\eq{
\hat p= \hat p(\hat T,\hat \mu^*; ~\tilde A, \tilde B)~,~~~~~\hat n= \hat n(\hat T,\hat \mu^*;~ \tilde A, \tilde B)~,
}
where $\tilde A$ and $\tilde B$ are given by Eq.~(\ref{AB}).
Presenting  $\hat \mu^*$ as $\hat \mu^*(\hat T, \hat n;~\tilde A, \tilde B) $ one obtains
\eq{
\hat p=\hat p(\hat T,\hat n;~\tilde A, \tilde B)~.
}
At fixed temperature $\hat T>\hat T_{\rm c}$ the system isotherm is a monotonic function of $\hat n$, i.e., $(\partial \hat p/\partial  \hat n)_{\hat T} >0$.  When $\hat T$  decreases, one comes to the critical isotherm $\hat T_{\rm c}$ for which  $(\partial \hat p/\partial  \hat n)_{\hat T_{\rm c}} =0$ at the critical density $\hat n_{\rm c}$. From these critical quantities one can then find the values of
$T_{\rm c}/T_{\rm c}^0$ and 
$n_{\rm c}/n_{\rm c}^0$ as functions of the two parameters, namely, the functions of $\tilde A$ and $\tilde B$. Therefore, the type of the CP, either Boltzmann CP or Bose CP, is defined by the values of the dimensionless parameters
$\tilde A$ and $\tilde B$. As shown in Appendix \ref{B}, in the non-relativistic approximation this  is reduced to a dependence on  the single parameter $\sigma = \sigma(\tilde A, \tilde B)$.

\section{Non relativistic  approximation}
\label{B}
The  spinodal is defined by the condition $(\partial p/\partial n)_T=0$, that can be written as
\eq{T + n\frac{d U}{d n}\omega_{\rm id}(T, \mu^*) = 0~.\label{spinodal}
}

\begin{figure}[t]
\includegraphics[width=0.48\textwidth]{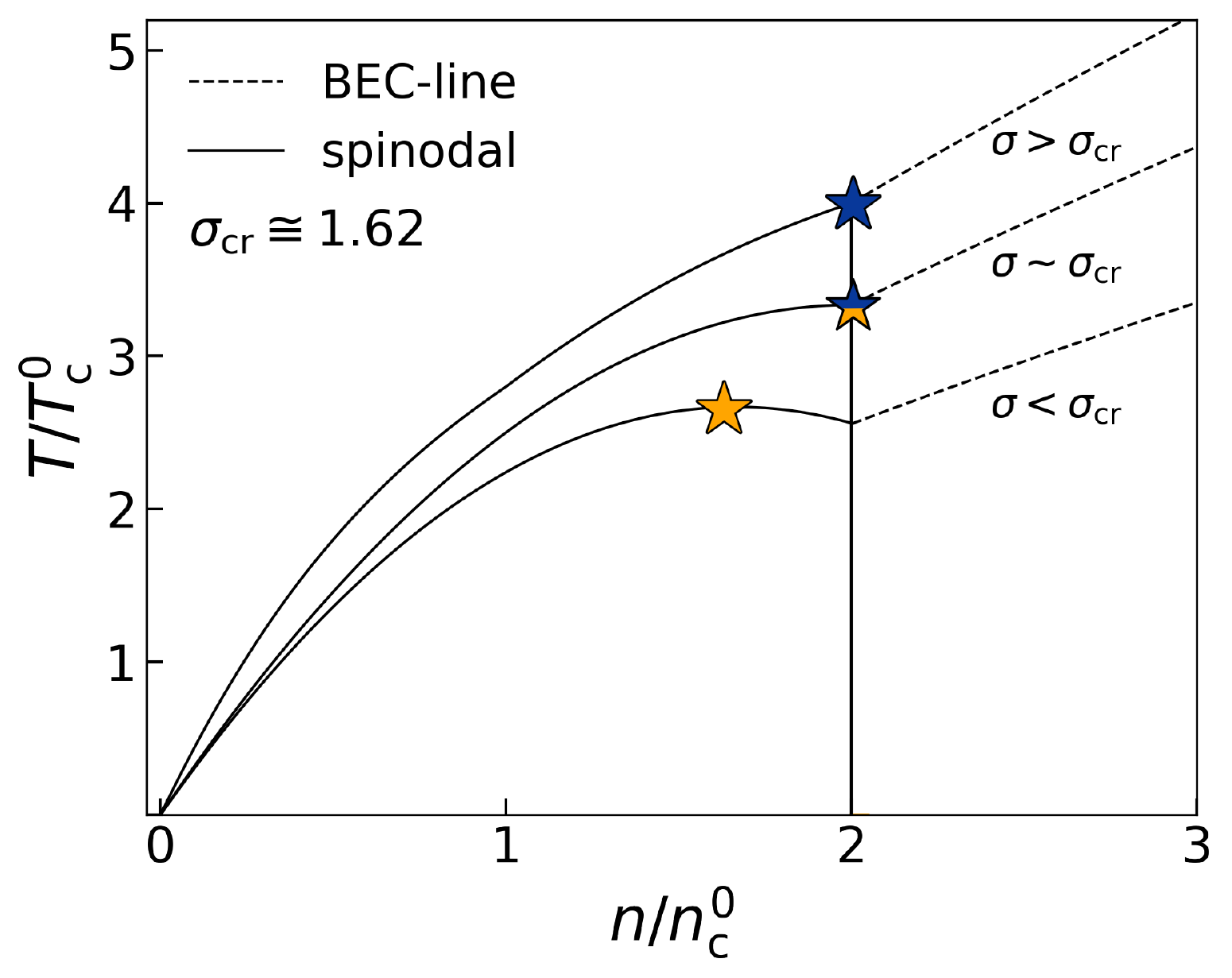}
\caption{The  spinodal lines $T_{\rm s}=T_{\rm s}(n)$ for different values of the parameter $\sigma$. The liquid spinodal  becomes the vertical line at $n = 2n^0_{\rm c}$.
} 
\label{fig7}
\end{figure}

The functions $T_{\rm s}=T_{\rm s}(n)$ given by Eq.~(\ref{spinodal}) are shown in Fig.~\ref{fig7}.  These spinodal lines intersect with the BEC lines $T_{\rm BC}(n)$ at $n=n^{\rm lim}=2n_{\rm c}^0$ as  shown in Fig.~\ref{fig7}.
The analytical approximations  valid near the
BEC line can be used for $\omega_{\rm id}$ in Eq.~(\ref{spinodal}) 
(see Refs.~\cite{Begun:2008hq, Begun:2006gj}):

\eq{\omega_{\rm id}(T,\mu^*) &= \frac{g m^{3/2}T^2}{2\sqrt{2}\pi n}(m - \mu^*)^{-1/2}~, \label{omega-id-approx}
\\
m - \mu^*(T) &= \frac{9\zeta^2(3/2)}{16\pi}\frac{(T - T_{\rm BC})^2}{T_{\rm BC}}~.\label{m-mu}
}
The BEC temperature $T_{\rm BC}$ in the non-relativistic approximation (\ref{TBC-1})
can be written as
\eq{\label{TBC-red}
\tilde T_{\rm BC} =\frac{T_{\rm BC}}{T^0_{\rm c}}
= 2\left[\frac{2\tilde n \sigma}{\zeta(3/2)}  \right]^{2/3}~,
}

where $\sigma$ is given by Eq.~(\ref{sigma}).

By substituting Eqs.~(\ref{omega-id-approx}-\ref{TBC-red}) into Eq.~(\ref{spinodal}) and rewriting it in the reduced variables, one finds 
the function $\tilde T_{\rm s}(\tilde n)$
at $\tilde{n}\rightarrow 2-0$.
The Boltzmann CP corresponds to 
$d\tilde{T}_{\rm s}/d\tilde{n} <0$  
and the Bose CP to $d\tilde{T}_{\rm s}/d\tilde{n} >0$
at $\tilde n=2$ (see Fig.~\ref{fig7}).
Thus, the condition
\eq{\label{sigma-cr}
\left.\frac{d\tilde T_{\rm s}}{d\tilde n}\right|_{\tilde n=2} ~=~0
}
corresponds to the transition between the Boltzmann CP and the Bose CP.

The solution of Eq.~(\ref{sigma-cr}) gives the critical value  for the parameter $\sigma=\sigma_{\rm cr}$. The straightforward analytical solution  of Eq.~(\ref{sigma-cr})
gives
\eq{\sigma_{\rm cr} ~=~ \frac{2\pi^{3/2}}{\zeta^2(3/2)}~\cong ~1.62~.
\label{sigma-162}
}

\section{Binodals near the Bose CP}
\label{C}
In a case of the Bose CP we introduce
\eq{
 \frac{n_{\rm c} - n_{\rm g}}{n_{\rm c}} & \equiv
\frac{\delta n_{\rm g}}{n_{\rm c}}= 
C_{\rm g}~\frac{T_{\rm c}-T}{T_{\rm c}}~,
\label{ng}\\
\frac{n_{\rm l} - n_{\rm c}}{n_{\rm c}} & \equiv 
\frac{\delta n_{\rm l}}{n_{\rm c}}=  
C_{\rm l}~\frac{T_{\rm c}-T}{T_{\rm c}}~\label{nl}.
}
At $T< T_{\rm c}-0$ one finds
$\mu_{\rm g}^*(T) <  m$ on the gaseous binodal and $\mu_{\rm l}^*=m$ on the liquid one.

At $T\rightarrow T_{\rm c}$ one has $\mu_{\rm g}^*\rightarrow m$ and, using Eq.~(\ref{m-mu}),  obtains the following system for dimensional binodal slope parameters $C_{\rm g}$ and $C_{\rm l}$:

\eq{\begin{split}
&2\left(C_g^3+C_l^3\right)=3\left(C_g^2-C_l^2\right)\frac{T_{\rm c}}{n_{\rm c}} \left. \frac{d n_{\rm id}(T,m)}{d T} \right|_{T = T_{\rm c}} \\&+2R(C_g^2-C_l^2)^{3/2}~,\end{split} \label{C-1}
\\&C_g=\frac{T_{\rm c}}{n_{\rm c}}\left. \frac{d n_{\rm id}(T,m)}{d T} \right|_{T = T_{\rm c}} + R(C_g^2-C_l^2)^{1/2}~,\label{C-2}
}
where $R = \sqrt{B}gT_cm^{3/2}/(4\sqrt{2}\pi)$.

In the non relativistic approximation the derivative will be
\eq{ \label{nid-nr}
\left.\frac{ 
dn_{\rm id}(T,m)}{dT} \right|_{T=T_{\rm c}} = \frac{3n_{\rm c}}{2T_{\rm c}}~.
}
Using Eqs.~\eqref{m-mu}, \eqref{C-1} and \eqref{C-2} one finds
\eq{\begin{split}
&(C_g^3+C_l^3) = \frac{9}{4}(C_g^2-C_l^2) \\&+ \left(\frac{\sigma_{cr}}{\sigma}\right)^{1/3}(C_g^2-C_l^2)^{3/2}~, \label{C1}
\end{split}
\\&C_g = \frac{3}{2}+ \left(\frac{\sigma_{cr}}{\sigma}\right)^{1/3}\sqrt{C_g^2-C_l^2}~. \label{C2}
}
From Eqs.~(\ref{C1}) and (\ref{C2}) 
at $\sigma \to \sigma_{\rm cr}+0$
\eq{C_g \to \infty, ~~~~C_l \to \infty, ~~~~ \frac{C_g}{C_l} \to \infty~,}
and

\eq{C_g = \frac{3}{2},~~~~ C_l = \frac{3}{4}, ~~~~ \frac{C_g}{C_l} = 2~, \label{ClCg-nr}}
at $\sigma \gg \sigma_{\rm cr}$.
Thus, there is an asymmetry between the gaseous and liquid binodals. This asymmetry becomes very large when $\sigma$ approaches its critical value $\sigma_{\rm cr}$.

At $\sigma=\sigma_{\rm cr}$ Eq.~(\ref{ng}) 
is broken and should be substituted by $\delta n_{\rm g}/n_{\rm c}=C_{\rm g}\sqrt{(T_{\rm c}-T)/T_{\rm c}}$. It means that for $\sigma=\sigma_{\rm cr}$ the critical exponent $\beta$ has the Boltzmann value $\beta=1/2$ at $n<n_{\rm c}=2n_{\rm c}^0$ and the Bose value $\beta =1$ at $n>n_{\rm c}=2n_{\rm c}^0$

With increasing $T_c$ value in  Eqs.~\eqref{C-1} and \eqref{C2} one obtains
a limiting behavior
\eq{\label{ClCg-large}
C_{\rm g}\cong \frac{T_{\rm c}}{n_{\rm c}}\, \left.\frac{dn_{\rm id}(T,m)}{dT} \right|_{T=T_{\rm c}}~,~~~~
C_{\rm l}~=~\frac{1}{2}\, C_{\rm g}~
}
that corresponds to the large $\sigma$ behavior (\ref{ClCg-nr}) in the non relativistic approximation.

In the ultra relativistic limit one has 
\eq{
\left.\frac{ 
dn_{\rm id}(T,m)}{dT} \right|_{T=T_{\rm c}} = \frac{3n_{\rm c}}{T_{\rm c}}~
}
instead of Eq.~(\ref{nid-nr}).
Using Eq.~(\ref{ClCg-large}) in the ultra-relativistic limit,
\eq{C_g = 3, ~~~~C_l = \frac{3}{2}, ~~~~\frac{C_g}{C_l} = 2~.}

\bibliography{references1.bib}
\end{document}